\documentclass[conference]{IEEEtran}

\usepackage{comment}
\usepackage{graphicx}
\usepackage{amsmath}
\usepackage{booktabs}
\usepackage{graphicx}
\usepackage{caption}
\usepackage{subcaption}
\usepackage{hyperref}
\usepackage{tikz}
\usetikzlibrary{positioning}
\usepackage{pgfplots}

\usepackage{xcolor}
\usepackage{soul}
\usepackage{multirow}

\usepackage{listings}
\usepackage{xcolor}

\lstdefinelanguage{json}{
  basicstyle=\scriptsize\ttfamily,
  numbers=none,
  showstringspaces=false,
  breaklines=true,
  frame=single,
  stringstyle=\color{red!70!black},
  keywordstyle=\color{blue},
  morekeywords={operator,c_function,rtl_module,interface,inputs,outputs,
                name,valid,ready,scheduling,latency,initiation_interval},
}

\def\BibTeX{{\rm B\kern-.05em{\sc i\kern-.025em b}\kern-.08em
    T\kern-.1667em\lower.7ex\hbox{E}\kern-.125emX}}

\title{
Programming Domain-Specific FPGA Hardblocks from HLS: An RTL Blackbox Approach
}

\author{
\IEEEauthorblockN{Ruthwik Reddy Sunketa}
\IEEEauthorblockA{Arizona State University\\
Tempe, AZ, USA\\
rsunketa@asu.edu}
\and
\IEEEauthorblockN{Jeevesh Choudhury}
\IEEEauthorblockA{Arizona State University\\
Tempe, AZ, USA\\
jchoudh3@asu.edu}
\and
\IEEEauthorblockN{Aman Arora}
\IEEEauthorblockA{Arizona State University\\
Tempe, AZ, USA\\
aman.kbm@asu.edu}
}

\IEEEoverridecommandlockouts
\IEEEpubid{}

\begin{document}

\maketitle

\begin{abstract}

Domain-specific Field Programmable Gate Array (FPGA) architectures increasingly integrate specialized hardblocks, such as Tensor Slices, to accelerate artificial intelligence and machine learning workloads. Despite their efficiency benefits, these architectures remain difficult to program, as designers typically rely on manual Register-Transfer Level (RTL) integration to access these hardblocks.
This paper presents a compiler-agnostic methodology that enables high-level synthesis (HLS) tools to target custom FPGA hardblocks directly from C/C++ code. Architectural hardblocks are exposed as schedulable C-level operators using an RTL blackbox abstraction with explicit latency and initiation-interval contracts, allowing the HLS scheduler to optimize around specialized hardware without manual RTL orchestration.
Unlike prior uses of HLS blackboxes for vendor IP integration, our approach treats blackboxes as architectural abstractions, enabling the scalable composition of C-level operators that can target custom FPGA hardblocks without compiler modification.

We evaluate the proposed flow using a Tensor Slice–based FPGA architecture with AMD Vitis HLS \cite{amd_vitis_hls} and the Verilog-to-Routing (VTR) toolchain \cite{Elgammal:2025:TRETS:VTR-9}.
Results across multiple matrix sizes show that designs generated using the proposed C-Blackbox flow achieve lower area–delay product than behavioral HLS baselines
while achieving substantially higher productivity-adjusted efficiency relative to handwritten RTL.
These results demonstrate that domain-specific FPGA architectures can be made accessible through HLS without sacrificing performance.

\end{abstract}

\section{Introduction}
\label{sec:intro}

\begin{figure}[t]
    \centering
    \includegraphics[width=\linewidth]{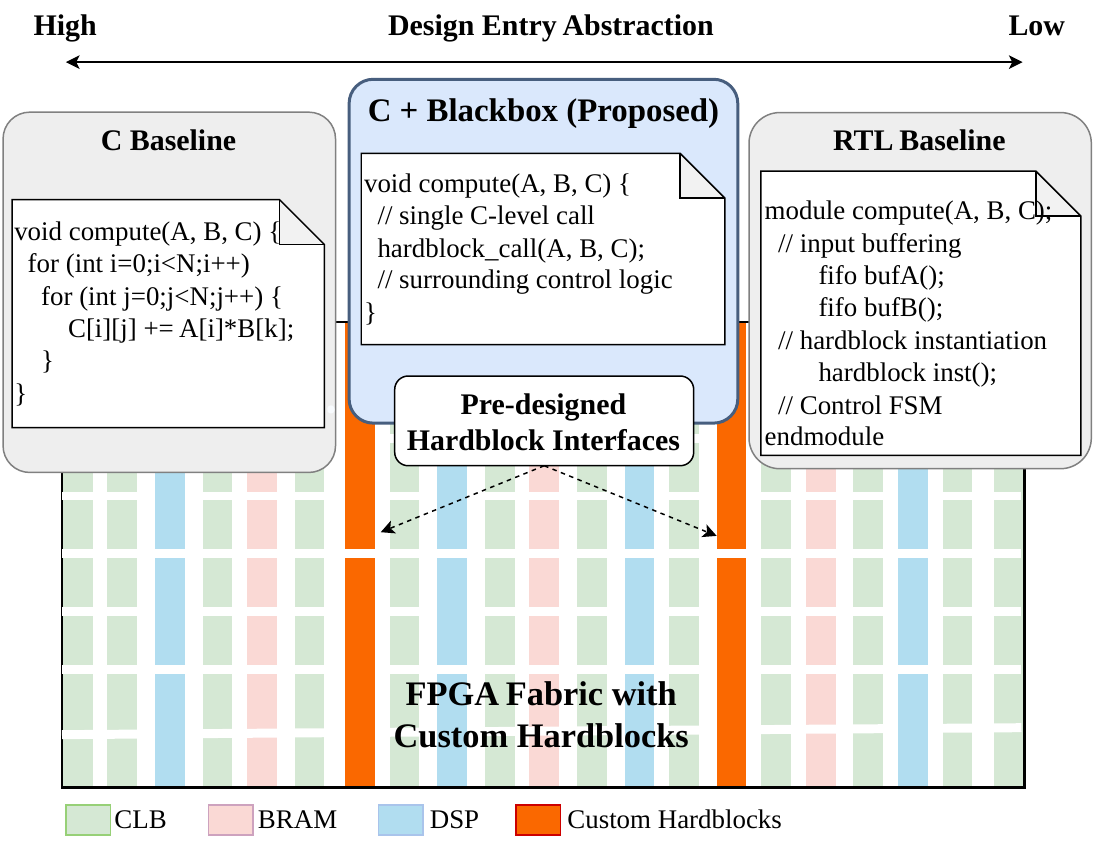}
    \caption{Comparison of three FPGA design flows targeting the same heterogeneous FPGA fabric. C-Baseline expresses computation entirely in high-level code. RTL-Baseline instantiates hardblocks at the RTL level. The C-Blackbox flow (proposed) integrates pre-designed RTL wrappers as blackbox operators within C code, bridging design productivity and hardware efficiency.}
    \label{fig:design_flows}
\end{figure}

Modern FPGAs are increasingly evolving from homogeneous fabrics of look-up tables (LUTs) into domain-specific architectures that integrate specialized hardblocks. To meet the performance and energy-efficiency demands of artificial intelligence (AI) and machine learning (ML) workloads, recent FPGA architectures incorporate coarse-grained primitives such as Tensor Slices~\cite{arora:tensor_slice:TRETS, taka:sparse_tensor:ISPGA} and in-memory computing units~\cite{arora:comefa:trets, wang:CCB:FCCM2021, chen:M4BRAM:FPT2023}. These architectural hardblocks offer substantial gains in throughput and area efficiency, but they remain difficult to program using existing design flows.

Currently, exploiting a new hardblock typically requires manual Register-Transfer Level (RTL) integration. Designers must explicitly instantiate hardblocks, manage cycle-accurate control, and coordinate low-level hardware interfaces, resulting in long development cycles and limited portability. Although HLS improves productivity by allowing designs to be expressed in C or C++, existing HLS tools cannot natively target custom architectural hardblocks. In the absence of architectural awareness, HLS compilers default to generic soft-logic implementations, underutilizing specialized hardware.
One possible approach to addressing this gap is to use compiler-centric approaches that modify the HLS compiler to explicitly support new architectural primitives.
Such methods create a tighter coupling between architectural evolution and compiler development. As a result, architectural exploration and compiler modification may become intertwined, complicating rapid evaluation and iteration on emerging FPGA hardblocks.

To address these challenges, we use the blackbox mechanism of HLS tools that supports integrating external RTL modules into a C design. 
Rather than using blackboxes solely as a means of incorporating external IP, our work elevates blackboxes to an architectural abstraction that defines how domain-specific hardblocks are programmed from HLS.

Our methodology exposes custom FPGA hardblocks to HLS tools as schedulable C-level operators. Rather than modifying the compiler, we represent custom hardblocks as blackbox operators with explicit latency and initiation-interval (II) specifications.
This abstraction allows the HLS scheduler to reason about the timing behavior of specialized hardware while remaining agnostic to its internal micro-architecture, eliminating the need for manual RTL system integration.

Figure \ref{fig:design_flows} illustrates the different design entry methods for heterogeneous FPGAs. The C Baseline (left) allows for fast development but relies entirely on the HLS tool to infer general-purpose logic 
in traditional FPGA fabrics (containing Configurable Logic Blocks (CLBs), DSPs, and BRAMs); it does not utilize specialized hardblocks, which results in lower performance and hardware efficiency.
The RTL Baseline (right) provides high hardware efficiency by manually instantiating hardblocks, but it requires the designer to explicitly implement complex control logic in RTL, which limits productivity.
Our proposed C + Blackbox flow provides a middle ground; it uses pre-designed RTL wrappers and metadata to expose specialized hardware as schedulable C-level operators. This allows the HLS tool to automatically manage system-level scheduling and control logic while achieving higher Quality-of-Results (QoR), facilitated by specialized hardblocks.

Compared to traditional RTL-based integration, which can require weeks of development and verification effort, our approach enables rapid deployment of hardware-optimized functionality directly from C/C++ code.
The proposed methodology decouples architectural innovation from compiler development by exposing custom FPGA hardblocks as schedulable C-level operators without modifying the HLS compiler.
In addition to increasing design productivity, this also enables architecture evaluation of specialized FPGA fabrics without manual RTL integration or invasive toolchain changes.

The contributions of this work are as follows:
\begin{itemize}    
    
    \item Our methodology introduces an architectural abstraction that uses the HLS blackbox mechanism as a programming model for domain-specific FPGA architectures.
    
    \item We represent FPGA hardblocks as blackbox operators in HLS,  allowing designers to develop large-scale systems from C/C++, thereby increasing design productivity.

    \item We demonstrate an end-to-end flow from C-design to FPGA implementation on a Tensor Slice-based fabric using AMD Vitis HLS and VTR toolchains.

    \item We quantitatively evaluate the productivity--QoR tradeoffs of the proposed approach using Tensor Slices as a case study, showing order-of-magnitude reductions in design effort while maintaining predictable performance and competitive productivity-adjusted efficiency.
    
\end{itemize}

Although we evaluate the proposed methodology using Tensor Slices and AMD Vitis HLS, the blackbox-based abstraction itself is general and can be applied to a wide range of domain-specific FPGA hardblocks and HLS toolchains that support external RTL integration.

\section{Background and Related Work} \label{sec:related_work}

\subsection{Domain-Specialized FPGA Architectures}

Recent FPGA architecture research has increasingly shifted toward \emph{domain-specific fabrics} that integrate specialized hardblocks to improve Deep Learning (DL) inference efficiency. Representative examples include tensor-oriented compute blocks, such as the DL-optimized Tensor Slices introduced by Arora~\emph{et~al.} \cite{arora:tensor_slice:TRETS}, which augment FPGAs with fixed-function low-precision systolic matrix multiplication units to substantially increase INT8 and FP16 compute density, as well as subsequent extensions that incorporate sparse tensor support \cite{taka:sparse_tensor:ISPGA}. 
Several works have explored compute-enabled BRAM architectures that integrate arithmetic capabilities directly into on-chip memories, including CCB \cite{wang:CCB:FCCM2021}, Compute-RAM \cite{arora2021computeRAM}, CoMeFa \cite{arora:comefa:trets}, M4BRAM \cite{chen:M4BRAM:FPT2023}, and  BRAMAC \cite{chen:bramac:2023}. 
All these architectures establish a clear design trend: embedding domain-specific, coarse-grained compute primitives into the FPGA fabric to improve hardware efficiency for DL workloads. As surveyed by Boutros~\emph{et~al.}~\cite{Butros:DL_fpga_arch_survey:IEEE}, these heterogeneous hardblock architectures demonstrate clear gains in throughput and energy efficiency and are commonly evaluated using research CAD frameworks such as VTR \cite{Elgammal:2025:TRETS:VTR-9}. However, despite their architectural diversity,  these hardblocks are used via manual RTL integration or model-specific hardware generators \cite{HPIPE:stan:fpt2022}, leaving programming such specialized FPGA architectures from general-purpose HLS flows as a significant open challenge.

\subsection{Tensor Slice Architecture}
As a representative case study for utilizing custom hardware from high-level languages, we target the Tensor Slice architecture \cite{arora:tensor_slice:TRETS} optimized for high-throughput matrix operations. Unlike traditional DSP blocks, the Tensor Slice is a coarse-grained primitive capable of performing $8 \times 8$ matrix-vector and matrix-matrix multiplications.

The Tensor Slice features a systolic array of 16 processing elements capable of dynamically supporting multiple precisions (including INT8, INT16, FP16, and BF16) and various tensor operations such as matrix-matrix and matrix-vector multiplications. To achieve high throughput, the architecture relies on explicitly controlled, cycle-by-cycle data orchestration and dedicated hardware chaining to scale computations across multiple physical blocks. While 
 significant gains in compute density and routing efficiency were demonstrated by the authors, the evaluation relied entirely on designing custom control logic and manually instantiating these hardblocks using Verilog \cite{arora:tensor_slice:TRETS, taka:sparse_tensor:ISPGA}. This leaves a critical gap in high-level programmability, which our work addresses by introducing an HLS methodology to seamlessly abstract, schedule, and compose these complex hardblocks using standard C/C++ workflows.

\subsection{High-Level Synthesis (HLS) Toolchains}

High-level synthesis tools translate high-level specifications (typically C or C++) into hardware implementations. While various research-oriented frameworks such as LegUp \cite{canis2013LegUp}, BAMBU \cite{ferrandi2021bambu}, CIRCT \cite{circt}, and Dynamatic \cite{dynamatic:isfpga:2018} exist, this work focuses on leveraging the mature scheduling and optimization capabilities of a commercial HLS tool (AMD Vitis HLS \cite{amd_vitis_hls}) to integrate custom FPGA hardblocks.

While the AMD Vitis HLS tool natively supports the integration of external RTL modules using blackbox mechanisms, such features are traditionally used for incorporating fixed, external IP blocks in an FPGA design. 
Our approach uses the blackbox mechanism to define and expose the architectural hardblocks intrinsic to the FPGA fabric.
This capability enables high-level programming of specialized hardware primitives without requiring manual RTL system integration.
\section{Proposed Flow} \label{sec:proposed_fow}

\begin{figure}[t]
    \centering
    \includegraphics[width=\linewidth]{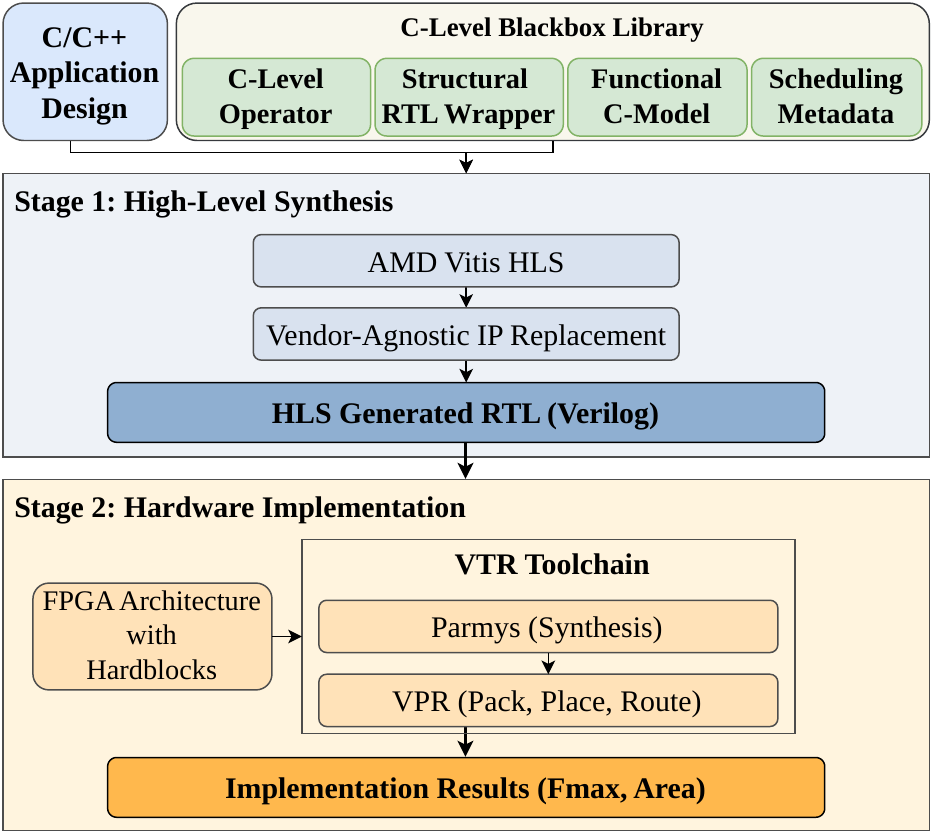}
    \caption{End-to-end design flow illustrating the integration of the proposed C-level blackbox API library with a commercial HLS tool (AMD Vitis HLS) and the VTR backend.}
    \label{fig:hls-vtr-pipeline}
\end{figure}

The core of our methodology is a design flow that bridges the abstraction gap between high-level C/C++ software specifications and specialized FPGA fabrics. As shown in Fig.~\ref{fig:hls-vtr-pipeline}, the flow enables a direct path from software-level descriptions to hardware implementations on custom FPGA architectures, without requiring low-level RTL integration.
We first describe the various components of this flow and then describe the end-to-end flow in this section.

\subsection{Flow Components}

The goal of the proposed flow is to expose hardblocks as C-level operators that can be scheduled by the HLS compiler, without requiring any compiler modification. To achieve this, each architectural hardblock is represented using a C-level function, a synthesizable RTL wrapper, a functional C model, and explicit scheduling metadata.

\subsubsection{C-level Operator}

From the perspective of the HLS compiler, hardblocks are treated as opaque blackboxes with well-defined timing behavior, represented as a C-level operator.
This operator allows micro-architectural complexity of the block to be hidden behind a programming interface.

A single physical hardblock may be represented by multiple distinct C-level functions or operators. For example, an underlying Tensor Slice hardblock may support both INT8 and FP16 matrix multiplication, which are exposed to the programmer as two separate C-level operators.
The compiler sees a collection of independent callable operators, which carry their own interface definition and scheduling metadata, even though all variants map to the same hardblock on the FPGA.

\begin{figure}[t]
\centering
\begin{lstlisting}[language=Verilog]
module ts_wrapper(clk, rst_n,
        rowA_data,rowA_valid, rowA_ready,
        colB_data,colB_valid, colB_ready,
        rowC_data,rowC_valid, rowC_ready);
  
  // Interface buffers (e.g., AXI-Stream <-> internal)
  fifo inst_A (...);
  fifo inst_B (...);
  fifo inst_C (...);

  // Tensor Slice hardblock instantiation
  tensor_slice_hw inst (
    .clk(clk), .rst_n(rst_n),
    .A_row(rowA_data), .B_col(colB_data),
    .C_row(hw_rowC)
  );

  // Control FSM: protocol translation 
  always @(posedge clk) begin
    case (state)
      IDLE:    // AXI-stream handshake
      START:   // streams -> hardblock inputs
      PROCESS: // hardblock output -> stream
    endcase
  end
endmodule
\end{lstlisting}
\caption{Representative structural RTL wrapper showing hardblock instantiation, protocol
translation (e.g., AXI-stream to Tensor Slice interface), and control logic.}
\label{fig:rtl_wrapper}
\end{figure}

\subsubsection{Structural RTL Wrapper}

The structural RTL wrapper provides the hardware realization of each C-level blackbox operator. As shown in Figure \ref{fig:rtl_wrapper}, the wrapper instantiates the physical hardblock and implements the control logic and interface adaptation required to align the HLS execution model with the hardblock’s native protocol.
The HLS-generated logic never directly interacts with the raw hardblock interface. All cycle-accurate control and hardware-specific timing is encapsulated within the wrapper, ensuring compatibility with the target hardblock.

\subsubsection{Functional C-Model}
Each blackbox operator's model includes a functional C-model that mirrors the behavior of its corresponding RTL wrapper. The C-model enables software-level simulation and early functional verification while preserving the productivity benefits of high-level design.
To ensure consistency between simulation and synthesis, the C-model must use the same data types and interfaces
as the RTL wrapper, including arbitrary-precision types and streaming constructs where appropriate.

\begin{figure}[t]
\centering
\begin{lstlisting}[language=json]
{
  "operator": "ts_wrapper",
  "c_function": "ts_wrapper",
  "rtl_module": "ts_wrapper",

  "interface": {
   "inputs": [ 
    { "name":"rowA_data", "valid":"", "ready":"" },
    { "name":"colB_data", "valid": "", "ready":"" } ],
   "outputs": [
      { "name":"rowC_data", "valid":"", "ready":"" } ]
  },

  "scheduling": {
    "latency": 2,
    "initiation_interval": 1
  }
}
\end{lstlisting}
\caption{Scheduling metadata for the \texttt{ts\_wrapper} blackbox operator, specifying
the ready/valid interface and the latency/II specifications used by the HLS scheduler.}
\label{figs:ts_metadata}
\end{figure}

\subsubsection{Scheduling Metadata}

Metadata forms the contract between the blackbox abstraction and the HLS compiler. Figure \ref{figs:ts_metadata} shows an example of the metadata (interface, latency, and II) of the hardware implementation for a C-level operator. This metadata allows the compiler to schedule operations, overlap independent invocations, and avoid structural hazards within the hardblock.
By decoupling scheduling information from implementation details, the flow enables the HLS tool to optimize surrounding logic while preserving predictable performance characteristics. This mechanism is central to achieving competitive QoR without requiring the compiler to be aware of hardblock internals.

\subsubsection{Scaling and Composition}
For problem sizes that exceed the capacity of a single hardblock instance, our flow supports scaling within the RTL wrapper. For example, a wrapper implementing $16 \times 16$ matrix multiplication may internally contain four $8 \times 8$ Tensor Slice instances, managing data movement and synchronization across the array.
A single wrapper can instantiate multiple hardblocks internally using tiling, chaining, or systolic-style interconnection with data distribution, synchronization, and result aggregation all being handled within the wrapper logic.
Crucially, this scaling is completely hidden from the HLS compiler. Regardless of the internal composition, the compiler continues to view the operation as a single functional unit with given latency and II.
Although this adds a one-time cost of building the abstraction models and wrappers for all operators, it enables architectural scalability without increasing C-level code complexity and allows large designs to be constructed from reusable 
primitives.

Additionally, the proposed flow also supports hierarchical scaling through composition at the C level. Larger designs can be constructed by invoking multiple instances of blackbox operators within the C program, allowing the HLS tool to schedule and pipeline these operator calls while relying on wrapper-level logic to manage local data movement and control. This approach enables problem sizes that exceed the capacity of a single blackbox operator to be expressed naturally at the C level, without exposing low-level hardblock interfaces or cycle-accurate control to the programmer.
This two-level scaling model, with wrapper-level composition for moderate sizes and C-level operator composition for larger designs, provides flexibility for balancing hardware efficiency and design complexity.

\subsection{End-to-End Flow}

The end-to-end design flow converts a C-level description of an application into a physical FPGA implementation using AMD Vitis HLS and VTR toolchain.
The flow requires two main inputs: the C application to be mapped to the target FPGA (containing C function calls), and a blackbox library that contains: (1) C header files to declare the C functions for each operator, (2) JSON metadata files specifying the interface, latency and II for each operator's hardware implementation, and (3) RTL wrappers to implement the blackbox operators.
This library creation is a one-time cost paid by flow developers. Users only need to write the C application embedded with C operator function calls.

The flow goes through two stages. In the first stage, Vitis HLS performs high-level synthesis to generate Verilog code that instantiates the RTL wrappers of the C operators.
During HLS, the tool treats each operator as a macro with the provided timing properties. The tool uses the metadata to organize the system control logic and generates a Verilog implementation that integrates the provided RTL wrappers. 
Commercial HLS tools, such as Vitis HLS, often infer vendor-locked IP blocks for standard operations, such as floating-point arithmetic or DSP functions. We replace these blocks with generic, synthesizable implementations compatible with open-source toolchains, ensuring the final design remains independent of vendor-specific hardware.

The resulting RTL is then mapped to the target FPGA using the Verilog-to-Routing (VTR) framework. To ensure compatibility, the hardblocks in the blackbox wrappers are designed to match the hardware model defined in the VTR architecture. The VTR backend performs synthesis, packing, placement, and routing, and provides QoR metrics such as frequency and area.
\section{Methodology} \label{sec:methodology}

To evaluate the proposed framework, we conduct a comparative study using various sizes of General Matrix Multiplication (GEMM) benchmarks across three design flows:
\begin{itemize}
    \item \textbf{C-Baseline}: A behavioral C implementation in which all computation is expressed at the C level and fully synthesized by Vitis HLS using standard operators that are mapped to traditional FPGA fabric (including CLBs, DSPs, and BRAMs).
    \item \textbf{C-Blackbox}: A C-level implementation that invokes Tensor Slice operations through the proposed blackbox API library. Specialized computation is mapped to custom hardblocks via RTL wrappers, while the surrounding control logic is synthesized by the HLS tool. This is the proposed flow.
    \item \textbf{RTL-Baseline}: A manually written RTL implementation that explicitly instantiates Tensor Slice modules and associated control logic. These designs use AXI-based interfaces from \cite{forencich_verilog_axi}.
\end{itemize}

The goal of this evaluation is not to establish that one design flow produces universally superior hardware implementations compared to another. In principle, highly optimized designs can be achieved using any of these flows given sufficient manual effort and domain expertise. Instead, our objective is to assess whether the proposed HLS-based flow can achieve competitive QoR while significantly reducing programmer effort.
The results should therefore be understood as a productivity--QoR tradeoff study, rather than a head-to-head competition between fully optimized design flows.

\subsection{Target FPGA Architecture}

All experiments target a heterogeneous FPGA architecture described in~\cite{arora:tensor_slice:TRETS}. The architecture consists of a conventional fabric of CLBs, DSPs, and BRAMs, augmented with columns of specialized Tensor Slice hardblocks.
Each CLB contains 10 6-input LUTs (which can be fractured into two 5-input LUTs), along with dedicated carry chains between CLBs. BRAMs have a capacity of 20 Kilobits and support true and simple dual-port modes with varied heights and widths (e.g., 512 × 40, 1024 × 20, and 2048 × 10). DSPs support multiple precisions, including 9 × 9, 18 × 19, 27 × 27, fp16, bf16, and fp32. The routing fabric consists of length 4 and length 16 wires with a channel width of 300.
This organization reflects emerging domain-specific FPGA architectures, where coarse-grained accelerators are embedded alongside general-purpose logic to improve performance and efficiency for compute-intensive workloads.

The Tensor Slice hardblock exposes a fixed-latency and explicitly controlled interface. 
For example, when configured for the 8x8 INT8 matrix multiplication mode, the Tensor Slice strictly consumes one row of the first operand matrix and one column of the second operand matrix per clock cycle. It eventually streams out one row of the result matrix per cycle, with an initiation interval (II) of 1 and a fixed latency of 24 cycles. 
These precise timing and I/O behavior 
can be seamlessly encoded as scheduling metadata for the HLS compiler, making it representative of the class of domain-specific primitives targeted by the proposed flow.

\begin{table*}[t]
\centering
\caption{Comparison of scaling behavior across matrix sizes using C-Baseline, Proposed, and RTL-Baseline design flows. ADP = Area $\times$ Latency (MWTA$\cdot$s).}
\label{tab:scaling_grouped}
\begin{tabular}{c c
                r r r r
                r r r r r}
\toprule
\textbf{GEMM size} &
\textbf{Design} &
\textbf{CLB} &
\textbf{DSP} &
\textbf{BRAM} &
\textbf{Tensor} &
\textbf{Total Area} &
\textbf{Latency} &
\textbf{ADP} &
\textbf{Efficiency} &
\textbf{Eff. per LoC.} \\
 &  &
\textbf{[\#]} &
\textbf{[\#]} &
\textbf{[\#]} &
\textbf{Slice [\#]} &
\textbf{[MWTA]} &
\textbf{[s]} &
\textbf{[MWTA$\cdot$s]} &
\textbf{[GMAC/s/MWTA]} &
\textbf{} \\
\midrule

\multirow{3}{*}{8$\times$8}
 & C-Baseline   & 1,463 & 16  & 5   & 0  & 4.59e+07 & 1.14e-06 & 5.233e+01 & 9.78e-09 & 9.14e-11 \\
 & Proposed   & 1,257 & 0   & 12  & 1  & 3.82e+07 & 9.41e-07 & 3.595e+01 & 1.42e-08 & 1.21e-10 \\
 & RTL-Baseline &   683 & 0   & 16  & 1  & 2.27e+07 & 7.75e-07 & 1.759e+01 & 2.91e-08 & 1.90e-11 \\

\cmidrule(lr){1-11}

\multirow{3}{*}{16$\times$16}
 & C-Baseline   & 1,982 & 64  & 166 & 0  & 9.57e+07 & 4.56e-06 & 4.364e+02 & 9.39e-09 & 5.83e-11 \\
 & Proposed   & 1,459 & 0   & 18  & 4  & 4.89e+07 & 3.68e-06 & 1.800e+02 & 2.27e-08 & 1.91e-10 \\
 & RTL-Baseline &   861 & 0   & 29  & 4  & 3.37e+07 & 3.13e-06 & 1.055e+02 & 3.88e-08 & 2.44e-11 \\

\cmidrule(lr){1-11}

\multirow{3}{*}{32$\times$32}
 & C-Baseline   & 4,692 & 256 & 307 & 0  & 2.43e+08 & 1.97e-05 & 4.787e+03 & 1.17e-08 & 5.29e-11 \\
 & Proposed   & 2,417 & 0   & 31  & 16 & 9.46e+07 & 1.62e-05 & 1.533e+03 & 2.14e-08 & 1.81e-10 \\
 & RTL-Baseline & 1,213 & 0   & 55  & 16 & 6.42e+07 & 1.42e-05 & 9.116e+02 & 3.60e-08 & 2.13e-11 \\

\bottomrule
\vspace{0.1cm}
\end{tabular}
\end{table*}

\subsection{HLS Configuration}

All HLS experiments are performed using AMD Vitis HLS v2024.1. Designs are synthesized from C++ to Verilog RTL targeting a Xilinx Artix-7 FPGA device (xc7a100t-csg324-1) using a uniform target clock frequency of 200 MHz with default optimization settings.
Each design undergoes C-level simulation for functional verification, followed by C synthesis to export synthesizable Verilog RTL. C/RTL co-simulation is performed using behavioral Verilog models of the Tensor Slice to validate correctness across the C-level abstraction and the corresponding RTL implementation.
The specified FPGA part is used solely to configure the HLS front-end and does not affect backend mapping, which uses the VTR toolchain.

\subsection{Evaluation Metrics}
\label{sec:eval-metrics}

We evaluate the proposed framework using a combination of conventional hardware QoR metrics and productivity-aware measures. Hardware efficiency is characterized by area (reported in Minimum-Width Transistor Area, MWTA), maximum operating frequency, end-to-end latency, steady-state throughput (measured in GMAC/s), and throughput per unit area. Design effort is quantified using user-written lines of code (LoC), excluding reusable library components.
To capture productivity--quality tradeoffs, we additionally report a productivity-adjusted efficiency metric defined as throughput per unit area normalized by LoC. This composite metric is used solely for comparative analysis across design methodologies and is intended to highlight how effectively hardware performance scales relative to programmer effort.

\section{Results}

\subsection{Design Productivity} 
We first evaluate the impact of the proposed RTL blackbox methodology on design productivity. LoC measures the amount of user-written design logic required to implement each kernel and excludes reusable library components, such as RTL wrappers for C-Blackbox and AXI interfaces for RTL-Baseline.
For the  $32\times32$ INT8 GEMM kernel, the C-Baseline implementation requires approximately 197 
LoC,  while the proposed C-Blackbox implementation reduces this to 118 lines by encapsulating architectural scaling and hardblock control within reusable RTL wrappers. 
In contrast, the RTL-Baseline requires 1,692 lines of Verilog, dominated by explicit control logic and hardblock coordination.

Overall, for the $32\times32$ INT8 design, the proposed blackbox-based flow achieves a  ~14$\times$ reduction in user-written code compared to handwritten RTL, directly translating to shorter development cycles and reduced verification effort. A similar trend is observed across other matrix sizes, with the blackbox approach consistently requiring the fewest LoC, followed by the C-Baseline, while RTL implementations require the most.

\begin{table*}[t]
\centering
\caption{Comparison of wrapper-level and C-level composition for a $32\times32$ GEMM kernel. Efficiency improvement is reported relative to the C-Baseline.}
\label{tab:native_vs_tiled}
\begin{tabular}{l
                r r r r
                r r r r r}
\toprule
\textbf{Design} &
\textbf{CLB} &
\textbf{DSP} &
\textbf{BRAM} &
\textbf{Tensor} &
\textbf{Total Area} &
\textbf{Latency} &
\textbf{ADP} &
\textbf{Efficiency} &
\textbf{Eff. Improv.} \\
 &
\textbf{[\#]} &
\textbf{[\#]} &
\textbf{[\#]} &
\textbf{Slice [\#]} &
\textbf{[MWTA]} &
\textbf{[s]} &
\textbf{[MWTA$\cdot$s]} &
\textbf{[GMAC/s/MWTA]} &
\textbf{($\times$ over C-Baseline)} \\
\midrule
Wrapper-level composition
& 2{,}417 & 0 & 31 & 16
& 9.46e+07 & 1.62e-05 & 1.53e+03 & 2.14e-08 & 1.83 \\

C-level composition
& 2{,}650 & 3 & 208 & 16
& 1.26e+08 & 1.74e-05 & 2.20e+03 & 1.49e-08 & 1.27 \\

C-Baseline
& 4{,}692 & 256 & 307 & 0
& 2.43e+08 & 1.97e-05 & 4.79e+03 & 1.17e-08 & 1.00 \\
\bottomrule
\end{tabular}
\end{table*}

\subsection{Quality of Results Across GEMM Sizes}

\begin{figure}[t]
    \centering
    \includegraphics[width=\linewidth]{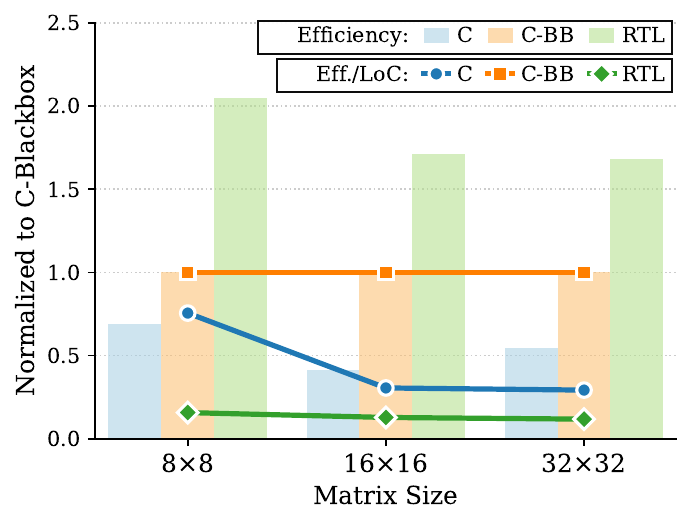}
    \caption{Comparison of throughput efficiency (bars) and efficiency per line of code (lines) across matrix sizes, normalized to the C-Blackbox implementation.}
    \label{fig:eff_plot}
\end{figure}

We next examine the quality-of-results tradeoffs introduced by the proposed abstraction across GEMM sizes, with particular attention to how hardware efficiency compares against fully handwritten RTL implementations. Table~\ref{tab:scaling_grouped} summarizes the scaling behavior of the three design flows across three GEMM matrix sizes (8×8, 16×16, and 32×32).

Across all matrix sizes, the RTL-Baseline consistently achieves the lowest total area and the highest raw throughput efficiency (GMAC/s/MWTA), reflecting the benefits of fully manual hardblock integration and tightly optimized, cycle-accurate control logic. As shown in Table~\ref{tab:scaling_grouped}, RTL designs establish an upper bound on achievable hardware efficiency for the target architecture.
The proposed C-Blackbox flow incurs a moderate reduction in raw throughput efficiency relative to RTL, typically within a factor of 1.5–2× across all evaluated matrix sizes, due to wrapper-level control logic, interface adaptation, and automated HLS scheduling. Nevertheless, C-Blackbox designs consistently outperform the C-Baseline, which relies entirely on HLS-generated soft logic and does not exploit Tensor Slice hardblocks.

Latency scales proportionally with matrix size across all flows, with no evidence of superlinear growth. Importantly, C-Blackbox designs maintain latency within 15–20\% of the RTL-Baseline across all configurations, demonstrating that the explicit latency and initiation-interval (II) metadata provided to the HLS scheduler is sufficient to preserve predictable timing behavior.

Despite lower peak efficiency than RTL, the C-Blackbox approach demonstrates significantly superior productivity-adjusted efficiency. As shown in Fig.~\ref{fig:eff_plot}, when throughput efficiency is normalized by user-written lines of code, the C-Blackbox flow exceeds the C-Baseline by up to 3× and the RTL-Baseline by approximately 7–9× across matrix sizes. Overall, these results show that the proposed C-Blackbox flow intentionally trades peak hardware efficiency for dramatically reduced design effort, yielding predictable performance and substantially higher productivity-adjusted efficiency than both behavioral HLS and handwritten RTL approaches.

\subsection{C-level composition vs RTL wrapper-level composition}

An important question is how hierarchical composition using smaller blackbox operators in C compares to composing a larger operator within an RTL wrapper. We evaluate this tradeoff by comparing wrapper-level composition, where the $32\times32$ GEMM kernel is implemented as a single blackbox operator with an RTL wrapper instantiating a $4\times4$ grid of Tensor Slice hardblocks and handling all control internally, to C-level composition, where the same $32\times32$ kernel is constructed by composing multiple $16\times16$ blackbox operators directly in C and relying on HLS-generated control logic.

Table~\ref{tab:native_vs_tiled} shows a representative comparison for the $32\times32$ GEMM kernel. Wrapper-level composition (C-Blackbox) achieves the highest throughput efficiency among the HLS-based designs, improving efficiency by 1.83× over the C-Baseline. 
This advantage is partly due to architectural features of the target FPGA: Tensor Slice hardblocks natively support chaining to form larger matrix-multiplication units, which is directly exploited by the RTL wrapper. In the current implementation, these low-level features are not exposed to HLS-level blackbox operators, preventing C-level composition from leveraging native Tensor Slice chaining and resulting in additional control overhead.

C-level composition still improves efficiency by 1.27× relative to the C-Baseline and offers greater flexibility by composing operators directly in C. 
Overall, these results show that wrapper-level composition is more effective for achieving high hardware efficiency, while C-level composition provides a reasonable tradeoff between efficiency and programmability. Both approaches deliver predictable performance without unexpected overheads.

\subsection{Summary of Findings}

Across all evaluated configurations, the proposed blackbox-based HLS flow consistently achieves competitive hardware QoR while reducing user-written code by an order of magnitude. No superlinear area growth, severe frequency degradation, or compounding abstraction overheads are observed as designs scale in size. Taken together, these results validate the proposed approach as a practical productivity--QoR tradeoff, enabling efficient access to domain-specific FPGA architectures without the burden of manual RTL integration or HLS compiler modification.

\section{Conclusion}

This work demonstrates that domain-specific FPGA architectures with custom hardblocks can be made accessible through  HLS tools without modifying the compiler or relying on manual RTL integration.
We present an end-to-end design flow from C designs to physical implementation for a Tensor Slice-based FPGA using AMD Vitis HLS and VTR toolchains.

Several directions remain for future exploration, such as automating the generation of C models, RTL wrappers, and scheduling metadata, which could further reduce integration effort and improve portability across architectures.
In addition, future work includes exposing low-level architectural features, such as native Tensor Slice chaining, to HLS-level operators to reduce control overhead and improve the efficiency of C-level composition.

\section{Acknowledgements}
This work was supported in part by National Science Foundation (grant number 2417658). Any opinions, findings, conclusions, or recommendations are those of the authors and not of the funding institutions.
The authors acknowledge Research Computing at Arizona State University for providing access to the Sol Supercomputer, which contributed to the results reported in this paper \cite{Sol-HPC:ASU23}.

\bibliographystyle{IEEEtran}
\bibliography{refs}

\end{document}